\documentclass[aps,pra,reprint,floatfix,amsmath,amssymb,longbibliography]{revtex4-2}

\usepackage{bm}
\usepackage{graphicx}
\usepackage{amsthm}
\usepackage[colorlinks=true,allcolors=blue]{hyperref}

\allowdisplaybreaks[1]

\newcommand{\bra}[1]{\langle #1 |}
\newcommand{\ket}[1]{| #1 \rangle}
\newcommand{\ev}[1]{\langle #1 \rangle}

\begin{document}

\title{Level ordering in the linear $E \otimes e$ dynamic Jahn-Teller model}

\author{Naoya Iwahara}
\email[]{naoya.iwahara@gmail.com}
\affiliation{Graduate School of Engineering, Chiba University, 1-33 Yayoi-cho, Inage-ku, Chiba-shi, Chiba 263-8522, Japan}

\date{\today}

\begin{abstract}
The ground level of the linear $E\otimes e$ dynamic Jahn-Teller model has long been believed to be the doublet with vibronic angular momentum $j = \pm\tfrac12$.
We prove that, for arbitrary coupling strength, the $n$th level $E_{nj}$ in the subspace of fixed $j$ strictly increases with $|j|$ for every $n$,
$E_{n,1/2} < E_{n,3/2} < E_{n,5/2} < \cdots$; in particular, the ground level is always the $j = \pm\tfrac12$ doublet.
The proof uses only a unitary transformation that lowers $j$ by one, combined with the variational principle.
We also show that the levels with fixed $j$ are nondegenerate and give explicit lower and upper bounds on the gap $E_{0,j+1} - E_{0j}$.
The same Hamiltonian describes Rashba-coupled particles in harmonic traps and, in nuclear physics, the static nucleon in a charged scalar meson field.
\end{abstract}

\maketitle

The dynamic Jahn-Teller effect, characterized by quantum entanglement between degenerate electronic orbitals and degenerate lattice vibrations, gives rise to unconventional phenomena in molecules, impurities in crystals \cite{Englman1972, Bersuker1989}, and correlated compounds \cite{Chibotaru2005, Iwahara2013, Iwahara2023, Iwahara2024}.
It not only leaves characteristic fingerprints in spectroscopic signals \cite{Englman1972, Bersuker1989, Iwahara2024} but also controls macroscopic properties \cite{Chibotaru2005, Iwahara2013, Iwahara2023}.
Isomorphic or closely related models also arise in other quantum systems, such as Rydberg atoms in optical tweezer arrays~\cite{Magoni2023}, two-level systems in bimodal cavities~\cite{Larson2008}, trapped ions coupled to their two transverse vibrational modes~\cite{Porras2012}, spin-orbit-coupled cold atoms in harmonic traps~\cite{Larson2009, Sinha2011, Hu2012PRL}, and semiconductor quantum dots~\cite{Kuan2004}.
The same mathematics appeared even earlier in nuclear physics: a nucleon fixed in space that emits and absorbs charged mesons, treated at strong coupling before the Jahn-Teller effect was formulated \cite{Wentzel1940}, is an $E\otimes e$ problem.

The linear $E\otimes e$ dynamic Jahn-Teller problem \cite{Moffitt1957a, Moffitt1957b, Longuet-Higgins1958}, an orbital doublet coupled to a doubly
degenerate vibration, is the simplest prototype capturing the nontrivial features \cite{Englman1972, Bersuker1989}.
The model has an O(2) symmetry \cite{Bersuker1989, Longuet-Higgins1958}, and the Hamiltonian conserves a half-integer vibronic angular momentum $j$ \cite{Longuet-Higgins1958}.
The model is only quasi-exactly solvable: isolated exact solutions exist at special couplings~\cite{Judd1979}, and otherwise the levels are obtained as roots of convergent series~\cite{Reik1987}, as for the Rabi model~\cite{Braak2011}; no closed form of the spectrum is known.
Various theoretical analyses and numerical calculations support the conjecture that the ground state belongs to $j = \pm\tfrac12$ and has the same symmetry as the electronic state \cite{Moffitt1957a, Moffitt1957b, Longuet-Higgins1958, Muramatsu1978, Ham1987, Sato2005, Requist2017, Ribeiro2018}.

In this Letter, we prove this conjecture, extend it to the ordering of all vibronic levels, and bound the gap between the lowest levels.

\begin{figure}[t]
\includegraphics[width=\columnwidth]{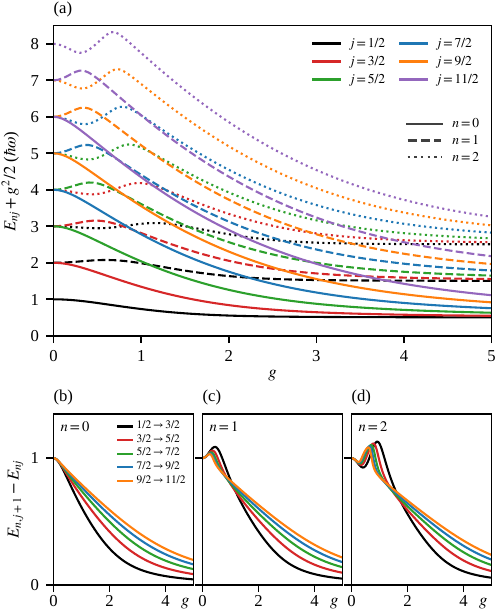}
\vspace{-4pt}
\caption{Vibronic levels of the linear $E\otimes e$ model (in units of $\hslash \omega$).
(a) The energy levels as functions of $g$.
$E_{nj} + g^2/2$ for $n = 0$ (solid), $n = 1$ (dashed) and $n = 2$
(dotted), $j = \tfrac12, \dots, \tfrac{11}{2}$ (colors).
$E_{nj}$ increases with $|j|$.
The energy gaps $E_{n,j+1}-E_{nj}$ for (b) $n=0$, (c) $n=1$, and (d) $n=2$.
All gaps are positive.
We calculated the levels numerically following Ref.~\cite{Muramatsu1978}.
}
\label{fig_levels}
\end{figure}

\textit{Model.}---
The model Hamiltonian consists of the harmonic oscillator Hamiltonian of the doubly degenerate modes and the vibronic coupling between the orbital doublet and the vibrations:
\begin{equation}
 H =
 \hslash \omega
 \left[
 \frac{1}{2}\left(p_\theta^2 + p_\varepsilon^2
   + q_\theta^2 + q_\varepsilon^2\right)
   + g \left(-q_\theta \sigma_z + q_\varepsilon \sigma_x\right)
   \right].
 \label{eq_Hcart}
\end{equation}
Here, $q_\theta$, $q_\varepsilon$ are the dimensionless normal coordinates of the $e$ mode with frequency $\omega$,
and $p_\theta$, $p_\varepsilon$ are their conjugate momenta.
$\sigma_x$, $\sigma_y$, $\sigma_z$ are the Pauli matrices acting on the electronic doublet $\{\ket{\theta}, \ket{\varepsilon}\}$,
and $g$ is the dimensionless vibronic coupling constant.
The normal coordinates and the conjugate momenta satisfy the following commutation relations:
\begin{align}
 [q_\gamma, q_{\gamma'}] =
 [p_\gamma, p_{\gamma'}] = 0, \quad
 [q_\gamma, p_{\gamma'}] = i\delta_{\gamma\gamma'},
 \label{eq_commun}
\end{align}
where $\gamma, \gamma' = \theta, \varepsilon$.

We can choose $g>0$ without loss of generality, because the simultaneous change in signs of
$g$, $q_\gamma$, and $p_\gamma$
leaves both $H$ \eqref{eq_Hcart} and the commutation relations \eqref{eq_commun} unchanged.
Hereafter, we set $\hslash \omega = 1$.

\textit{Symmetry of the model.}---
A direct calculation shows that the Hamiltonian \eqref{eq_Hcart} commutes with the vibronic angular momentum \cite{Longuet-Higgins1958} defined by
\begin{align}
 J = L - \tfrac12 \sigma_y,
 \quad
 L = q_\theta p_\varepsilon - q_\varepsilon p_\theta,
 \label{eq_J}
\end{align}
i.e., $[H, J] = 0$.
Hence, $H$ has an emergent continuous rotational symmetry SO(2) generated by $J$, and
$H$ and $J$ have simultaneous eigenkets labeled by the vibronic angular momentum $j$ together with an index $n$:
\begin{equation}
 H|\Psi_{nj}\rangle = E_{nj}|\Psi_{nj}\rangle, \quad J|\Psi_{nj}\rangle = j|\Psi_{nj}\rangle.
 \label{eq_labels}
\end{equation}
We denote by $\mathcal{H}_j$ the subspace with $J = j$.
We enumerate the levels so that $E_{nj} \le E_{n+1,j}$ for all $n$ ($= 0, 1, 2, \dots$) and $j$.
The eigenvalues of $J$ are the half-integers $j = \pm\tfrac12, \pm\tfrac32, \dots$ because the states must be single-valued in $(q_\theta, q_\varepsilon)$ space \cite{Longuet-Higgins1958}, as shown below.

All $E_{nj}$ are doubly degenerate due to the reflection symmetry \cite{Bersuker1989}.
The reflection, which changes $q_\varepsilon \to -q_\varepsilon$ together with $\ket{\varepsilon} \to -\ket{\varepsilon}$, is represented by
\begin{align}
 K = \sigma_z \exp(i \pi N_\varepsilon), \quad N_\varepsilon = \frac{1}{2}(p_\varepsilon^2 + q_\varepsilon^2 - 1).
 \label{eq_K}
\end{align}
A direct calculation shows that
\begin{align}
 [H, K] = 0, \quad \{J, K\} = 0.
 \label{eq_reflection}
\end{align}
Since $H(K|\Psi_{nj}\rangle) = KH|\Psi_{nj}\rangle = E_{nj}(K|\Psi_{nj}\rangle)$ and $J(K|\Psi_{nj}\rangle) = -KJ|\Psi_{nj}\rangle = -j (K|\Psi_{nj}\rangle)$, $K$ maps $\mathcal{H}_j$ onto $\mathcal{H}_{-j}$ preserving the energies, and
\begin{align}
 E_{nj} = E_{n,-j}.
 \label{eq_Enj}
\end{align}
Together with $K$, the SO(2) rotations form the group O(2).
Hereafter, we consider only positive $j$.

To see why $j$ is half-integer, and for the proofs below, we use polar coordinates, $q_\theta = \rho\cos\phi$, $q_\varepsilon = \rho\sin\phi$, and the complex electronic basis $\ket{\pm} = (\ket{\theta} \pm i\ket{\varepsilon})/\sqrt2$, i.e., the eigenvectors of $\sigma_y$ with eigenvalues $\pm 1$.
In these variables, the Hamiltonian becomes
\begin{align}
 H &= -\frac{1}{2}\left[
       \left(\frac{\partial^2}{\partial\rho^2} + \frac{1}{\rho}\frac{\partial}{\partial\rho}\right)
      - \frac{L^2}{\rho^2}\right]
      + \frac{\rho^2}{2} + g\rho\, M(\phi),
 \label{eq_Hpolar}
\end{align}
where $M$ and $L$ are, respectively,
\begin{align}
 M(\phi) &= -\left(e^{-i\phi}\ket{-}\bra{+} + e^{i\phi}\ket{+}\bra{-}\right),
 \label{eq_Mladder}
\\
 L &= -i\frac{\partial}{\partial\phi}.
 \label{eq_L}
\end{align}
With the same basis, the vibronic angular momentum is
\begin{align}
 J = L - \frac{1}{2}\tau_z,
 \label{eq_Jpolar}
\end{align}
where $\tau_z = |+\rangle\langle +| - |-\rangle \langle -|$.

The eigenkets of $J$ with eigenvalue $j$ have the form
\begin{equation}
 |\Phi_j(\rho,\phi)\rangle =
 \frac{1}{\sqrt{2\pi}}\left[
   e^{i(j+\frac12)\phi} f_+(\rho)\ket{+}
 + e^{i(j-\frac12)\phi} f_-(\rho)\ket{-}\right],
 \label{eq_sector}
\end{equation}
with the norm
\begin{equation}
 \ev{\Phi_j|\Phi_j} = \int_0^\infty \left(|f_+|^2 + |f_-|^2\right)\rho\, d\rho =1,
 \label{eq_norm}
\end{equation}
where the radial functions $f_\pm(\rho)$ are defined for $\rho \ge 0$.
Since $|\Phi_j(\rho, \phi)\rangle$ is invariant under $\phi \rightarrow \phi + 2\pi$, $l_\pm = j\pm \tfrac12$ must be integers; hence $j$ is half-integer.
Within $\mathcal{H}_j$, which consists of all $\ket{\Phi_j}$ \eqref{eq_sector} and which $H$ maps into itself, the vibrational angular momentum $L$ takes only the two values $l_\pm$.

\textit{Numerical vibronic levels.}---
Figure \ref{fig_levels}(a) shows the lowest levels $E_{nj}$ ($n = 0, 1, 2$; $j = \tfrac12, \dots, \tfrac{11}{2}$), obtained by numerical diagonalization of $H$ \eqref{eq_Hcart} \cite{Muramatsu1978}. Three features of the levels underlie the analysis below.
First, the energy levels of $H$ are discrete because the potential energy grows to infinity as $\rho \to \infty$ (Appendix~\ref{app_discrete}), and they are bounded from below by $-g^2/2$.
The lower bound follows because the potential energy $\rho^2/2 + g\rho M(\phi)$ has the eigenvalues $\rho^2/2 \pm g\rho \ge -g^2/2$ and the kinetic energy is non-negative, so that $H \ge -g^2/2$.
Second, the levels within each $\mathcal{H}_j$ never coincide, $E_{nj} < E_{n+1,j}$ [Fig.~\ref{fig_levels}(a)].
Levels with different $j$, in contrast, can cross; for example, $E_{1,1/2}$ and $E_{0,5/2}$ cross at $g \approx 1.09$.
Third, the $n$th level rises with $j$, $E_{n,j+1} - E_{nj} > 0$, for each $n$ and at every $g$ shown [Figs.~\ref{fig_levels}(b)-(d)].
These observations lead us to the following theorem.

\textit{Theorem.}---
For every real $g$, $n = 0, 1, 2, \dots$, and half-integer $j \ge \tfrac12$,
\begin{align}
 &\text{(I)} \qquad E_{nj} < E_{n+1,j},
 \label{eq_mainI}\\
 &\text{(II)} \qquad E_{nj} < E_{n,j+1}.
 \label{eq_main}
\end{align}
That is, (I) the levels in each subspace $\mathcal{H}_j$ are nondegenerate, and (II) $E_{n,1/2} < E_{n,3/2} < E_{n,5/2} < \cdots$; consequently, the ground state of $H$ is the $j = \pm\tfrac12$ doublet, which has the same symmetry as the electronic doublet.

\textit{Proof of (I).}---
We show that, in a suitable basis of $\mathcal{H}_j$, $H$ takes the form of an infinite Jacobi (tridiagonal) matrix with nonzero off-diagonal elements \cite{Moffitt1957a, Longuet-Higgins1958, Muramatsu1978}; the eigenvalues of such a matrix are simple \cite[Sec.~1.2, Remark~1.10]{TeschlJacobi}.
We expand the states of $\mathcal{H}_j$ in a vibronic basis of product states $\ket{N, L}\ket{\pm}$, where $\ket{N, L}$ are the eigenstates of the harmonic part of $H$, the two-dimensional oscillator $\frac12(p_\theta^2 + p_\varepsilon^2 + q_\theta^2 + q_\varepsilon^2)$, with $N = 0, 1, 2, \dots$ vibrational quanta (energy $N + 1$) and vibrational angular momentum $L = -N, -N+2, \dots, N$.
By Eq.~\eqref{eq_sector}, a product state in $\mathcal{H}_j$ has $L = l_+$ with $\ket{+}$ or $L = l_-$ with $\ket{-}$.
Since $L$ runs from $-N$ to $N$ in steps of two, $L$ is even for even $N$ and odd for odd $N$; of the two values $l_\pm$, which differ by one, exactly one is therefore possible for each $N \ge l_-$.
Writing $N = l_- + k$, the basis of $\mathcal{H}_j$ thus consists of $\ket{k} = \ket{N, l_-}\ket{-}$ for even $k$ and $\ket{k} = \ket{N, l_+}\ket{+}$ for odd $k$ ($k = 0, 1, 2, \dots$).
The coupling $g\rho M(\phi)$ contains $\rho e^{\pm i\phi} = q_\theta \pm i q_\varepsilon$, which changes $N$ by $\pm 1$; with suitable phases of the $\ket{k}$, $H$ is therefore tridiagonal,
\begin{equation}
 H\ket{k} = b_{k-1}\ket{k-1} + a_k\ket{k} + b_k\ket{k+1},
 \label{eq_jacobi}
\end{equation}
with $a_k = k + j + \tfrac12$, $b_k = g\sqrt{(k + 2j + 1)/2}$ for even $k$, $b_k = g\sqrt{(k + 1)/2}$ for odd $k$, and $b_{-1} = 0$.
For an eigenket $H\ket{\Psi} = E\ket{\Psi}$ in $\mathcal{H}_j$, the components $u_k = \langle k|\Psi\rangle$ obey $b_{k-1}u_{k-1} + (a_k - E)u_k + b_k u_{k+1} = 0$, because $E u_k = \bra{k}H\ket{\Psi}$ equals the inner product of $H\ket{k}$ with $\ket{\Psi}$ ($H$ is Hermitian and each $\ket{k}$ has finite energy).
For $g \ne 0$ all $b_k \ne 0$, so the recurrence fixes every $u_k$ once $u_0$ is given. Because the $\ket{k}$ form a complete orthonormal basis of $\mathcal{H}_j$, an eigenket with $u_0 = 0$ vanishes, and two eigenkets with the same energy in $\mathcal{H}_j$ are proportional to each other.
For $g = 0$ the levels $a_k$ are distinct.
Hence the levels in $\mathcal{H}_j$ are nondegenerate.
\hfill$\blacksquare$

\textit{Proof of (II) for $n=0$.}---
Let $U$ be the unitary operator defined in polar coordinates by
\begin{equation}
 U = e^{i\phi}.
 \label{eq_U}
\end{equation}
$U$ lowers $L$ \eqref{eq_L} and $J$ \eqref{eq_J} by one:
\begin{equation}
 U L U^\dagger = L - 1, \qquad U J U^\dagger = J - 1.
 \label{eq_ULU}
\end{equation}

Two consequences follow.

(i) \emph{$U^\dagger$ maps $\mathcal{H}_{j+1}$ onto $\mathcal{H}_j$.}
For $|\Phi_{j+1}\rangle \in \mathcal{H}_{j+1}$,
\begin{align}
 J\, U^\dagger|\Phi_{j+1}\rangle &= U^\dagger (U J U^\dagger)|\Phi_{j+1}\rangle
 \nonumber\\
 &= U^\dagger (J-1)|\Phi_{j+1}\rangle
 \nonumber\\
 &= j\,U^\dagger|\Phi_{j+1}\rangle,
\end{align}
and hence, $U^\dagger |\Phi_{j+1}\rangle \in \mathcal{H}_j$.
Since $U^\dagger$ is unitary and its inverse $U$ maps $\mathcal{H}_j$ back to $\mathcal{H}_{j+1}$,
\begin{equation}
 U^\dagger : \mathcal{H}_{j+1} \to \mathcal{H}_{j}
 \label{eq_mapU}
\end{equation}
is a norm-preserving bijection.

(ii) \emph{Transformation of the Hamiltonian.}
In Eq.~(\ref{eq_Hpolar}), only the centrifugal term contains $\partial/\partial\phi$.
Using $U L^2 U^\dagger = (U L U^\dagger)^2 = (L-1)^2 = L^2 - 2L + 1$,
\begin{equation}
 U H U^\dagger = H - \frac{2L - 1}{2\rho^2}.
 \label{eq_UHU}
\end{equation}

Equations~\eqref{eq_mapU} and \eqref{eq_UHU} relate the energy expectation values in $\mathcal{H}_j$ and $\mathcal{H}_{j+1}$.
For any $|\Phi_{j+1}\rangle \in \mathcal{H}_{j+1}$ and $|\tilde{\Phi}_j\rangle = U^\dagger |\Phi_{j+1}\rangle$,
\begin{equation}
  \langle \tilde{\Phi}_j|H|\tilde{\Phi}_j\rangle
  =
  \langle \Phi_{j+1}|H|\Phi_{j+1}\rangle -
  \left\langle \Phi_{j+1}\middle|\frac{2L-1}{2\rho^2}\middle|\Phi_{j+1}\right\rangle.
 \label{eq_Ediff}
\end{equation}
The expectation value of $(2L-1)/(2\rho^2)$ is positive, and here the two-valuedness of $L$ in $\mathcal{H}_{j+1}$ is essential.
Writing $|\Phi_{j+1}\rangle$ in the form of Eq.~\eqref{eq_sector} with $j \to j+1$, we have $L = j + \tfrac32$ on the $\ket{+}$ component and $L = j + \tfrac12$ on the $\ket{-}$ component, so that $(2L-1)/2 = j+1$ and $j$, respectively, which are both positive.
Hence,
\begin{align}
 \left\langle \Phi_{j+1}\middle|\frac{2L-1}{2\rho^2}\middle|\Phi_{j+1}\right\rangle
 &= \int_0^\infty \frac{(j+1)\,|f_+|^2 + j\,|f_-|^2}{\rho^2}\,\rho\,d\rho
 \nonumber\\
 &\ge j \int_0^\infty \frac{|f_+|^2 + |f_-|^2}{\rho^2}\,\rho\,d\rho
 \nonumber\\
 &= j\,\bra{\Phi_{j+1}}\rho^{-2}\ket{\Phi_{j+1}} > 0 .
 \label{eq_positive}
\end{align}
For the finiteness of $\bra{\Phi_{j+1}}\rho^{-2}\ket{\Phi_{j+1}}$, see Appendix~\ref{app_finite}.

Now we take $|\Phi_{j+1}\rangle$ in Eq.~\eqref{eq_Ediff} to be the normalized ground state $|\Psi_{0,j+1}\rangle$ in $\mathcal{H}_{j+1}$, so that $|\tilde{\Phi}_j\rangle = U^\dagger|\Psi_{0,j+1}\rangle$.
Writing $\ev{A}_{nj} = \bra{\Psi_{nj}}A\ket{\Psi_{nj}}$ for the expectation value of an operator $A$ in the eigenket $\ket{\Psi_{nj}}$, we obtain
\begin{align}
  \langle \tilde{\Phi}_j|H|\tilde{\Phi}_j\rangle
  =
  E_{0,j+1} - \left\langle\frac{2L-1}{2\rho^2}\right\rangle_{0,j+1}.
  \label{eq_trial}
\end{align}
Here, $|\tilde{\Phi}_j\rangle \in \mathcal{H}_j$ is not necessarily an energy eigenket.
The variational principle in $\mathcal{H}_j$ states that $E_{0j} \le \bra{\tilde{\Phi}_j}H\ket{\tilde{\Phi}_j}$ (Appendix~\ref{app_tech}).
Combining this with Eqs.~(\ref{eq_trial}) and (\ref{eq_positive}), we obtain
\begin{align}
 E_{0j} &\le \bra{\tilde{\Phi}_j} H \ket{\tilde{\Phi}_j}
 \nonumber\\
 &= E_{0,j+1} - \left\langle\frac{2L-1}{2\rho^2}\right\rangle_{0,j+1}
 \nonumber\\
 &\le E_{0,j+1} - j\ev{\rho^{-2}}_{0,j+1}
 \nonumber\\
 &< E_{0,j+1},
 \label{eq_chain}
\end{align}
which is the $n = 0$ case of Eq.~(\ref{eq_main}); its third line gives the lower bound in Eq.~(\ref{eq_bound}) below.
\hfill$\blacksquare$

\textit{Proof of (II) for $n>0$.}---
We now extend the result to the excited levels, using the min-max principle (Appendix~\ref{app_excited} gives an alternative proof based on the linear variational method, and Appendix~\ref{app_minmax} the rigorous conditions).
We build a normalized trial state for the $n$th level in $\mathcal{H}_j$ from the $n+1$ orthonormal states $U^\dagger\ket{\Psi_{k,j+1}}$ ($k = 0, 1, \dots, n$).
We choose the coefficients $c_k$ of the normalized trial state
\begin{equation}
 \ket{\tilde{\Theta}_j} = \sum_{k=0}^{n} c_k\, U^\dagger\ket{\Psi_{k,j+1}}
 \label{eq_Theta}
\end{equation}
so that $\ket{\tilde{\Theta}_j}$ is orthogonal to the eigenkets belonging to the $n$ lowest levels in $\mathcal{H}_j$:
\begin{equation}
  \langle \Psi_{mj}|\tilde{\Theta}_j\rangle = 0, \quad m = 0, 1, \dots, n-1.
 \label{eq_orth}
\end{equation}
These are $n$ homogeneous linear equations for the $n+1$ unknowns $c_k$, which always have a non-trivial solution.
For $n=0$ there is no condition, and $\ket{\tilde{\Theta}_j}$ reduces to the trial state $\ket{\tilde{\Phi}_j}$ of the proof for $n = 0$.

The energy expectation value $\bra{\tilde{\Theta}_j}H\ket{\tilde{\Theta}_j}$ can be bounded from below and from above.
Expanding $\ket{\tilde{\Theta}_j}$ \eqref{eq_Theta} in the eigenkets in $\mathcal{H}_j$,
we obtain $\ket{\tilde{\Theta}_j} = \sum_{m (\ge n)} d_m \ket{\Psi_{mj}}$ because of Eq.~\eqref{eq_orth}.
Hence,
\begin{equation}
 \bra{\tilde{\Theta}_j}H\ket{\tilde{\Theta}_j} = \sum_{m (\ge n)} E_{mj}|d_m|^2 \ge E_{nj} \sum_{m (\ge n)} |d_m|^2 = E_{nj}.
 \label{eq_mm}
\end{equation}
Since $\ket{\tilde{\Theta}_j} = U^\dagger\ket{\Theta_{j+1}}$ with $\ket{\Theta_{j+1}} = \sum_k c_k\ket{\Psi_{k,j+1}} \in \mathcal{H}_{j+1}$, Eqs.~\eqref{eq_Ediff} and \eqref{eq_positive} give
\begin{align}
 \bra{\tilde{\Theta}_j}H\ket{\tilde{\Theta}_j}
 &= \sum_{k=0}^{n} |c_k|^2 E_{k,j+1} - \left\langle \Theta_{j+1}\middle|\frac{2L-1}{2\rho^2}\middle|\Theta_{j+1}\right\rangle
 \nonumber\\
 &\le E_{n,j+1} - j\,\bra{\Theta_{j+1}}\rho^{-2}\ket{\Theta_{j+1}}
 \nonumber\\
 &< E_{n,j+1}.
 \label{eq_upper}
\end{align}
Combining Eqs.~\eqref{eq_mm} and \eqref{eq_upper}, we obtain $E_{nj} < E_{n,j+1}$.
This argument proves Eq.~(\ref{eq_main}) for all $n$ and all $j \ge \tfrac12$, giving
$E_{n,1/2} < E_{n,3/2} < E_{n,5/2} < \cdots$.
\hfill$\blacksquare$

\textit{Bounds on the gap.}---
The proof for $n = 0$ also bounds the gap between the lowest levels of neighboring subspaces from both sides:
\begin{equation}
 j\ev{\rho^{-2}}_{0,j+1} \le E_{0,j+1} - E_{0j} \le (j+1)\ev{\rho^{-2}}_{0j},
 \label{eq_bound}
\end{equation}
where the lower bound holds for $j \ge \tfrac12$ and the upper bound for $j \ge \tfrac32$.
The lower bound is the third line of Eq.~\eqref{eq_chain}.
The upper bound follows from the same transformation used in the opposite direction.
$U$ maps $\mathcal{H}_j$ onto $\mathcal{H}_{j+1}$, and $U^\dagger H U = H + (2L + 1)/(2\rho^2)$.
The variational principle with the trial state $U\ket{\Psi_{0j}} \in \mathcal{H}_{j+1}$ therefore gives $E_{0,j+1} \le E_{0j} + \ev{(2L+1)/(2\rho^2)}_{0j}$, where $(2L+1)/2 = j + 1$ and $j$ on the $\ket{+}$ and $\ket{-}$ components, respectively.
For $j \ge \tfrac32$ both components of $\ket{\Psi_{0j}}$ have $L \ge 1$, so $\ev{\rho^{-2}}_{0j}$ and the energy of the trial state are finite (Appendices~\ref{app_finite} and \ref{app_tech}, with $j+1$ replaced by $j$ and $l_s - 1$ by $l_s + 1$); for $j = \tfrac12$ the $\ket{-}$ component has $L = 0$, $\ev{\rho^{-2}}_{0,1/2}$ need not be finite, and the argument gives no upper bound.

The lower bound can be expressed explicitly in terms of $g$ and $j$.
Write $E = E_{0,j+1}$ and $R = \ev{\rho^2}_{0,j+1}^{1/2}$.
The Cauchy-Schwarz inequality gives $1 = \ev{\rho^{-1}\rho}_{0,j+1}^2 \le \ev{\rho^{-2}}_{0,j+1}R^2$ and $\ev{\rho}_{0,j+1} \le R$.
Since the kinetic energy is nonnegative and $\rho^2/2 + g\rho M(\phi) \ge \rho^2/2 - g\rho$, we have $E \ge R^2/2 - gR$, i.e., $R \le g + (g^2 + 2E)^{1/2}$.
Moreover, the variational principle with the trial state $\ket{k = 0}$ of $\mathcal{H}_{j+1}$ gives $E_{0,j+1} \le \bra{k=0}H\ket{k=0} = a_0 = j + \tfrac32$ [Eq.~\eqref{eq_jacobi} with $j \to j+1$].
Hence
\begin{align}
 E_{0,j+1} - E_{0j} &\ge j\left[g + \left(g^2 + 2E_{0,j+1}\right)^{1/2}\right]^{-2}
 \nonumber\\
 &\ge j\left[g + \left(g^2 + 2j + 3\right)^{1/2}\right]^{-2}.
 \label{eq_explicit}
\end{align}
At strong coupling the right-hand side approaches $j/(4g^2)$, which has the $g^{-2}$ dependence of the gap $(2j+1)/(2g^2)$ of the rotational levels in the trough \cite{Longuet-Higgins1958}.

\textit{Discussion.}---
We have proved the conjecture on the ground state of the linear $E \otimes e$ dynamic Jahn-Teller model and its extension to all excited levels.
The theorem is reminiscent of the ordering $E(n, l) < E(n, l+1)$ for a spinless particle in a scalar central potential, with $l$ the orbital angular momentum, which follows from the positivity of the centrifugal term~\cite[Sec. 2.2]{GrosseMartin1997}.
The present problem is, however, fundamentally different: the potential is a $2\times2$ matrix, and the nonadiabatic mixing of its two sheets is essential.

The proof uses only two properties of $H$:
(i) $H = L^2/(2\rho^2) + H'$, where $H'$ commutes with $e^{i\phi}$, which yields Eq.~\eqref{eq_UHU};
and (ii) $L \ge 1$ in $\mathcal{H}_{j+1}$ for $j \ge \tfrac12$, which guarantees Eq.~\eqref{eq_positive}.
The proof therefore remains valid if the harmonic potential is replaced by a radial potential $V(\rho)$ and $g\rho$ by a function $gF(\rho)$, provided that $V(\rho) - |gF(\rho)| \to \infty$ as $\rho \to \infty$.
With position and momentum interchanged, Eq.~\eqref{eq_Hcart} also describes an atom with Rashba spin-orbit coupling in an isotropic harmonic trap~\cite{Larson2009, Sinha2011} and an electron in a parabolic quantum dot with Rashba coupling at zero magnetic field~\cite{Kuan2004}; the theorem therefore also determines the ordering of their single-particle levels, with $j$ the $z$ component of the total angular momentum; this ordering had been obtained only numerically or asymptotically at strong coupling~\cite{Sinha2011, Hu2012PRL, Hu2012PRA}.

The argument requires the conservation of $J$ and property (ii).
It does not apply to the $E \otimes e$ problem with quadratic vibronic coupling, which breaks the SO(2) symmetry and hence the conservation of $J$; when strong, the quadratic coupling can even change the symmetry of the ground state~\cite{Koizumi1999}.
In the multimode $E\otimes e$ model with linear vibronic couplings, $J = \sum_i L_i - \tfrac12\sigma_y$, where $i = 1, 2, \dots$ labels the modes, is still conserved.
When all the modes have the same frequency, an orthogonal transformation of the modes concentrates the vibronic coupling in a single effective mode~\cite{OBrien1972}, and the problem reduces to the single-mode one plus free oscillators.
Since the ground state of the free oscillators has zero angular momentum, the ground state is again the $j = \pm\tfrac12$ doublet.
When the frequencies differ, no such reduction exists, and property (ii) fails: in a subspace of fixed $j$, the vibrational angular momentum $L_i$ of each mode takes all integer values, the excess in one mode being compensated by the others.
Hence the correction $(2L_i - 1)/(2\rho_i^2)$ produced by the transformation $e^{i\phi_i}$ of mode $i$ can be both positive and negative.
Thus the present approach does not determine the level ordering in this case, which remains open.

\textit{Vibronic reduction.}---
Because the ground level carries the symmetry $E$ of the electronic doublet at every coupling strength, electronic perturbations such as strain and Zeeman terms that are weak compared with the gap to the next level act within it as their electronic counterparts multiplied by vibronic reduction factors~\cite{Child1961, Ham1965}.

\textit{Nucleon in a meson field.}---
Equation~\eqref{eq_Hcart} also describes a nucleon fixed in space that emits and absorbs charged mesons~\cite{Wentzel1940}.
The proton and the neutron play the role of the electronic doublet, the two components of the meson field play the role of the $e$ modes~\cite{Slonczewski1963}, and $J + \tfrac12$ is the total electric charge $Q$.
If all mesons have the same energy, the theorem says that the proton and the neutron ($Q = 1$ and $0$) form the ground level at every coupling strength, and that the lowest levels with $Q = 2$ and $-1$, then $Q = 3$ and $-2$, and so on, lie progressively higher; Ref.~\cite{Wentzel1940} found this order at strong coupling.
Both consequences will be discussed in detail elsewhere.

In summary, the present result establishes, for all coupling strengths, the level structure that underlies the vibronic physics of $E\otimes e$ systems and of isomorphic systems such as Rashba-coupled cold atoms, quantum dots, and a nucleon in a charged meson field.

\appendix
\setcounter{secnumdepth}{1}

\section{Discreteness of the energy levels}
\label{app_discrete}

The energy levels of $H$ \eqref{eq_Hcart} are purely discrete because the potential grows to infinity.
Noting that the potential is equal to or larger than its lowest eigenvalue and that $(\rho/2-g)^2 \ge 0$ ($-g\rho \ge -\rho^2/4-g^2)$, the potential fulfills the following inequality: $\rho^2/2 + g\rho M(\phi) \ge \rho^2/2 - g\rho \ge \rho^2/4 - g^2$.
Thus, $H \ge H' \equiv \tfrac12 (p_\theta^2 + p_\varepsilon^2) + \tfrac14 (q_\theta^2 + q_\varepsilon^2) - g^2$.
The harmonic oscillator on the right-hand side has eigenvalues $\omega'(n + 1)-g^2$, where $\omega'=1/\sqrt{2}$ and $n = 0, 1, 2, ...$.
According to the min-max theorem, the $n$th lowest eigenvalues of $H$ and $H'$, denoted by $\mu_n$ and $\mu'_n$, satisfy $\mu_n \ge \mu_n'$ \cite[Cor.~4.13]{Teschl2014}.
Since $\mu_n' \rightarrow \infty$ as $n$ increases, so does $\mu_n$, which guarantees that the energy levels of $H$ are purely discrete \cite[Thm.~XIII.64]{ReedSimon4}.
The same argument applies within each $\mathcal{H}_j$.

\section{Finiteness of $\bra{\Phi_{j+1}}\rho^{-2}\ket{\Phi_{j+1}}$}
\label{app_finite}

Let $\ket{\Phi_{j+1}}$ be a state in $\mathcal{H}_{j+1}$ with finite energy.
Because $\rho^2/2 + g\rho M(\phi) \ge -g^2/2$, its kinetic energy obeys
\begin{align}
 T[\Phi_{j+1}] &= \sum_{s=\pm}\int_0^\infty
 \left[\frac12\left|\frac{df_s}{d\rho}\right|^2
 + \frac{l_s^2}{2\rho^2}|f_s|^2\right]\rho\,d\rho
 \nonumber\\
 &\le \bra{\Phi_{j+1}}H\ket{\Phi_{j+1}} + \frac{g^2}{2} .
\end{align}
Both terms in the square brackets are non-negative, and in $\mathcal{H}_{j+1}$ we have $l_\pm \ge j + \tfrac12 \ge 1$, hence
\begin{equation}
 \bra{\Phi_{j+1}}\rho^{-2}\ket{\Phi_{j+1}}
 \le \sum_{s=\pm} \int_0^\infty \frac{l_s^2}{\rho^2}|f_s|^2 \rho\,d\rho
 \le 2\,T[\Phi_{j+1}] < \infty .
 \label{eq_finite}
\end{equation}
For an eigenfunction this is also evident from its behavior at the origin:
$f_s(\rho) = \rho^{l_s} F_s(\rho^2)$ with $F_s$ smooth~\cite{Sato2005}, so
$|f_s|^2\rho^{-2}\rho \propto \rho^{2l_s - 1}$ is integrable at $\rho = 0$.

\section{Admissibility of the trial states}
\label{app_tech}

The variational principle holds for every normalized state with finite kinetic and potential energy
\cite[Thm.~2.14]{Teschl2014};
$H$ applied to the trial state need not even give a normalizable state.
For $\ket{\Phi_{j+1}} \in \mathcal{H}_{j+1}$ with finite energy, the potential energy of $\ket{\tilde{\Phi}_j} = U^\dagger\ket{\Phi_{j+1}}$ equals that of $\ket{\Phi_{j+1}}$, and its kinetic energy is obtained from $T[\Phi_{j+1}]$ (Appendix~\ref{app_finite}) by replacing $l_s \to l_s - 1$;
since $0 \le l_s - 1 < l_s$, $T[\tilde{\Phi}_j] \le T[\Phi_{j+1}] < \infty$.
For $j = \tfrac12$ the $\ket{-}$ component of $\ket{\tilde{\Phi}_j}$ has $L = 0$ and radial function $f_- \propto \rho$ near the origin.
This function has a kink at the origin, but its gradient is bounded, so its kinetic energy is finite.
A finite energy expectation value is all the variational principle requires; the trial state need not be smooth or satisfy the boundary conditions of the eigenfunctions.
A true eigenfunction in $\mathcal{H}_{1/2}$ has instead a finite, smooth $L = 0$ component at the origin~\cite{Sato2005}.

\section{Alternative proof for $n>0$}
\label{app_excited}
Here, we show an alternative proof of the theorem for $n>0$ by using the linear variational method \cite{Hylleraas1930, MacDonald1933}.
If $H$ is diagonalized within an $(n+1)$-dimensional subspace $\mathcal{W} \subset \mathcal{H}_j$, the resulting eigenvalues $\lambda_0 \le \lambda_1 \le \cdots \le \lambda_n$ are upper bounds to the exact levels in $\mathcal{H}_j$, $E_{kj} \le \lambda_k$ ($k = 0, 1, \dots, n$); this applies within $\mathcal{H}_j$ because $H$ maps $\mathcal{H}_j$ into itself.
In particular, the $n$th exact level in $\mathcal{H}_j$ lies below the largest energy expectation value in $\mathcal{W}$,
\begin{equation}
 E_{nj} \le \lambda_n = \max_{|\Phi\rangle \in \mathcal{W},\, \ev{\Phi|\Phi} = 1} \bra{\Phi}H\ket{\Phi} .
 \label{eq_HUM}
\end{equation}

Let $\mathcal{V}^{(n+1)}_{j+1}$ be the $(n+1)$-dimensional subspace formed by all linear combinations $|\Theta_{j+1}\rangle = \sum_{k=0}^n c_k \ket{\Psi_{k,j+1}}$ ($c_k \in \mathbb{C}$), where $\ket{\Psi_{k,j+1}}$ are the eigenkets in $\mathcal{H}_{j+1}$.
For a normalized $\ket{\Theta_{j+1}} \in \mathcal{V}^{(n+1)}_{j+1}$,
\begin{equation}
 \bra{\Theta_{j+1}}H\ket{\Theta_{j+1}} = \sum_{k=0}^n |c_k|^2 E_{k,j+1} \le E_{n,j+1}.
 \label{eq_Vn}
\end{equation}
Let $\mathcal{W} = U^\dagger\mathcal{V}^{(n+1)}_{j+1}$, which is an $(n+1)$-dimensional subspace of $\mathcal{H}_j$ by Eq.~\eqref{eq_mapU}, since $U^\dagger$ is unitary.
Every normalized element of $\mathcal{W}$ is $\ket{\tilde{\Theta}_{j}} = U^\dagger \ket{\Theta_{j+1}}$ with normalized $\ket{\Theta_{j+1}} \in \mathcal{V}^{(n+1)}_{j+1}$, and Eqs.~(\ref{eq_UHU}), (\ref{eq_positive}), and (\ref{eq_Vn}) give
\begin{align}
 \bra{\tilde{\Theta}_{j}}H\ket{\tilde{\Theta}_{j}}
 &= \bra{\Theta_{j+1}}H\ket{\Theta_{j+1}} - \left\langle \Theta_{j+1}\middle|\frac{2L-1}{2\rho^2}\middle|\Theta_{j+1}\right\rangle
 \nonumber\\
 &\le E_{n,j+1} - j\,\bra{\Theta_{j+1}}\rho^{-2}\ket{\Theta_{j+1}}
 \nonumber\\
 &< E_{n,j+1}.
 \label{eq_each}
\end{align}
Since $\mathcal{W}$ is finite dimensional, the maximum in Eq.~(\ref{eq_HUM}) is attained by some state of
$\mathcal{W}$, for which Eq.~(\ref{eq_each}) holds; hence $E_{nj} \le \lambda_n < E_{n,j+1}$.
\hfill$\blacksquare$

\section{Variational principle for excited states}
\label{app_minmax}
The statement used in the proof for $n>0$ is the upper-bound half of the min-max (Courant-Fischer-Weyl) principle.
Let $A$ be self-adjoint and bounded below with purely discrete eigenvalues, and let $\ket{\psi_1}, \dots, \ket{\psi_k}$ be linearly independent states of finite kinetic and potential energy \cite[Thm.~2.14]{Teschl2014}.
The trial states $\ket{\psi}$ need not be eigenfunctions, $A\ket{\psi_i}$ need not be normalizable, and the trial states need not satisfy the boundary conditions of the eigenfunctions.
If $\bra{\psi}A\ket{\psi} < \lambda\ev{\psi|\psi}$ for every nonzero linear combination $\ket{\psi}$ of the $\ket{\psi_i}$, then $A$ has at least $k$ eigenvalues, counted with multiplicity, below $\lambda$~\cite[Thm.~4.15(i)]{Teschl2014}.
With $k = n+1$ this is Eq.~(\ref{eq_HUM}), the Hylleraas-Undheim-MacDonald theorem~\cite{Hylleraas1930, MacDonald1933}.

In the proof for $n>0$ we apply it to $A = H$ restricted to $\mathcal{H}_j$, with the trial states $U^\dagger\ket{\Psi_{k,j+1}}$ (see Appendices~\ref{app_discrete} and \ref{app_tech} for the discreteness of the levels and the finite energy of the trial states).

\end{document}